\documentclass[aps,twocolumn,showpacs,amsmath,amssymb]{revtex4} 
 \usepackage{graphicx} 
 \usepackage[colorlinks=true, linkcolor=red, urlcolor=blue, citecolor=green]{hyperref}

 \def\T{\textstyle}
 \def\l{\left}
 \def\r{\right}
 \def\nf{n_{\!f}}

 \def\be{\begin{equation}}
 \def\ee{\end{equation}}
 \def\bea{\begin{eqnarray}}
 \def\eea{\end{eqnarray}}
 \def\bean{\begin{eqnarray*}}
 \def\eean{\end{eqnarray*}}
 \def\esim{\mathrel{\rlap{\lower0.2em\hbox{$-$}}\raise0.2em\hbox{$\sim$}}}
 \def\gsim{\mathrel{\rlap{\lower0.2em\hbox{$\sim$}}\raise0.2em\hbox{$>$}}}
 \def\ksim{\mathrel{\rlap{\lower0.2em\hbox{$\sim$}}\raise0.2em\hbox{$<$}}}
\begin{document}

\title{The QCD collisional energy loss revised}
\author{A.~Peshier}
\affiliation{
 Institut f\"{u}r Theoretische Physik,
 Universit\"{a}t Giessen, 35392 Giessen, Germany}
\date{\today}

\begin{abstract} \noindent
It is shown that to leading order the QCD collisional energy loss reads $dE/dx \sim \alpha(m_D^2)T^2$. Compared to prevalent expressions, $dE^B/dx \sim \alpha^2 T^2 \ln(ET/m_D^2)$, which could be considered adaptions of the (QED) Bethe-Bloch formula, the rectified result takes into account the running coupling, as dictated by renormalization. As one significant consequence, due to asymptotic freedom, the collisional energy loss becomes independent of the jet energy $E$.
Some implications with regard to heavy ion collisions are pointed out.
\end{abstract}


\maketitle

One of the key arguments for the creation of a `new state of matter' in heavy ion collisions at RHIC is the observed jet quenching \cite{jet quenching EXP}, which {\em inter alia} probes the jet's energy loss in the traversed matter.
In a quark gluon plasma (QGP) there are two effects causing a jet to lose energy: elastic collisions with deconfined partons \cite{Bjorken:1982tu, Braaten:1991jj}, or induced gluon radiation \cite{Wang:1991xy, Baier:1996kr, Gyulassy:2000fs}.
Presuming a dominance of the second mechanism, experimental findings have often been interpreted in terms of a purely radiative loss. However, the data-adjusted parameters (either $\hat q$ or $dN_g/dy$, depending on the approach) are found to be considerably larger than theoretically expected or even in conflict with a strong constraint from $dS/dy$ (see e.\,g.\ \cite{Muller:2005wi}) -- which calls for a collisional component of the energy loss.
The effect of collisions (as estimated within the framework \cite{Bjorken:1982tu, Braaten:1991jj}) might actually be larger than conceded for a long time \cite{Mustafa:2003vh, Wicks:2005gt}. The fact that such estimates depend crucially on the {\em assumed} value of the coupling should motivate us to scrutinize the principal question {\em `What is $\alpha$?'}.
Aside from its phenomenological relevance it will also lead to interesting theoretical insight.

\smallskip

Following Bjorken's intuitive considerations \cite{Bjorken:1982tu}, consider the propagation of a jet through a static QGP at a temperature $T \gg \Lambda$, where the coupling is small.
Its mean energy loss per length can be calculated from the rate of binary collisions with partons of the medium, as determined by the flux and the cross section,
\be
	\frac{dE_j}{dx}
	=
	\sum_s
	\int_{k^3} \rho_s(k)\, \Phi 
	\int dt\, \frac{d\sigma_{\!js}}{dt}\, \omega \, .
	\label{eq: dEdx start}
\ee
Here $\rho_s = d_s n_s$ is the density of scatterers, with $d_g = 16$ and $d_q = 12\nf$ being the gluon and quark degeneracies for $\nf$ light flavors, and $n_\pm(k) = \l[ \exp(k/T) \pm 1 \r]^{-1}$ in the ideal gas approximation.
Furthermore, $\Phi$ denotes a dimensionless flux factor, $t$ the 4-momentum transfer squared, and $\omega = E-E'$ the energy difference of the incoming and outgoing jet.
Focusing on the dominating scatterings with small momentum transfer, the cross sections read
\be
	\frac{d\sigma_{\!js}}{dt}
	=
	2\pi C_{\!js}\, \frac{\alpha^2}{t^2} \, ,
	\label{eq: dsigmadt}
\ee
with $C_{qq} = \frac49$, $C_{qg} = 1$, and $C_{gg} = \frac94$.
For $E$ and $E'$ much larger than the typical momentum of the thermal scatterers, $k \sim T$, the relation of $t$ to the angle $\theta$ between the jet and the scatterer simplifies to
\be
	t = -2(1-\cos\theta)k\omega \, ,
	\label{eq: t(eps)}
\ee
and the flux factor can be approximated by $\Phi = 1-\cos\theta$.

At this point, Bjorken integrated in Eq.~(\ref{eq: dEdx start})
\be
	\Phi\int_{t_1}^{t_2}\! dt\, \frac{d\sigma_{\!js}}{dt}\, \omega
	=
	\frac{\pi C_{\!js} \alpha^2}{-k}
	\int_{t_1}^{t_2} \frac{dt}t
	=
	\frac{\pi C_{\!js} \alpha^2}{k}\,
	\ln\frac{t_1}{t_2} \, ,
	\label{eq: Bjorkens t-int}
\ee
imposing both an IR and UV regularization. 
The soft cut-off is related to the Debye mass, $|t_2| = \mu^2 \sim m_D^2 \sim \alpha T^2$, describing the screening of the exchanged gluon in the medium.
The upper bound of $|t|$ was reasoned to be given by the maximum energy transfer: very hard transfers, say $\omega \approx E$, effectively do not contribute to the energy loss; in this case the energy is collinearly relocated to the scatterer.
Assuming $\omega_{max} = E/2$ implies $t_1 = -(1-\cos\theta)kE$, hence
$dE_j^B/dx = \pi\alpha^2 \sum_s C_{\!js}\int_{k^3} k^{-1} \rho_s \ln\l((1-\cos\theta)kE/\mu^2\r)$.
Replacing now, somewhat pragmatically, the logarithm by $\ln(2\langle k \rangle E/\mu^2)$ and setting $\langle k \rangle \rightarrow 2T$, Bjorken obtained
\be
	\frac{dE_{q,g}^B}{dx}
	=
	\l( \T\frac23 \r)^{\!\pm 1}
	2\pi \l( 1+\T\frac16 \nf \r) \alpha^2 T^2 \ln\frac{4TE}{\mu^2} \, ,
	\label{eq: dEdx Bjorken}
\ee
which differs for quark and gluon jets only by the prefactor.
This expression can be regarded as a relativistic adaption of the (QED) Bethe-Bloch formula \cite{Jackson}, which describes the ionization/excitation energy loss of charged particles in matter as determined by the scatterer density, and with the logarithm reflecting the kinematics and the long-range Coulomb-type interaction.

There are various refinements of Bjorken's `practical' calculation, aiming at the precise determination of the cut-offs. Worth accentuating is the approach of Braaten and Thoma \cite{Braaten:1991jj} who studied, within HTL perturbation theory, the propagation of a fermion through a QED plasma (and applied their method also to QCD).
For light quarks, the result reproduces the generic form (\ref{eq: dEdx Bjorken}), with $\mu \rightarrow m_D$ in the logarithm and the factor 4 replaced by some function of $\nf$ \footnote{As an aside, the HTL result is obtained	in
	Bjorken's approach (with a careful $k$-integration) with the IR cut-off 
	$\mu^2 = m_D^2/2$.}.

A remark concerning a pragmatic usage of such `Bjorken-type' formulae seems apposite here. Applied to experimentally relevant temperatures and rather low $E$, a {\em formally} resulting negative loss has been interpreted, at times, as an energy transfer of the thermal medium to the `soft jet'. This interpretation, however, is untenable in the given framework: the jet always loses energy in a collision, $\omega > 0$, cf.~(\ref{eq: t(eps)}). A negative result for $dE^B/dx$ is {\em de facto} the consequence of interchanged boundaries in the integration (\ref{eq: Bjorkens t-int}). Since $\mu^2$ is the minimal $|t|$ by definition, $dE^B/dx$ should instead be set to zero for $4E < \mu^2/T$. 
This concern, though, will prove irrelevant by the following considerations.

\smallskip

In Bjorken's derivation, $\alpha$ is a {\em fixed} parameter. 
Conceptionally, such a tree-level approximation may be appropriate for QED. The strong interaction, however, is known to vary considerably over the range of scales probed, e.\,g., in heavy ion collisions. Thus, in QCD one should study quantum corrections to the tree level processes, whose renormalization will specify unambiguously the value of `the' coupling.

For the sake of transparency of the argument, consider first the analogous case of electron scattering in massless QED. 
There are three types of (divergent) loop corrections to the $t$-channel tree-level process, see e.\,g.\ \cite{PeskinSchroeder}.
First, the exchanged photon is dressed by a self-energy. Then, encoded in the quantity $Z_1$, there are vertex corrections and finally, via the field strength renormalization $Z_2$, self-energy corrections to the external fermions. Yet, due to the identity $Z_1 \equiv Z_2$, in QED the renormalized coupling is determined only by the boson self-energy.

It is appropriate to renormalize the theory (i.e, to fix its parameters) by a scattering experiment at $T=0$.
The relevant part of the matrix element leading to the (vacuum) cross section corresponding to (\ref{eq: dsigmadt}) is $\alpha/\!\l( P^2-\Pi_{\rm vac}(P^2) \r)$. Here $\alpha$ denotes the {\em bare} coupling, and $\Pi_{\rm vac}$ is the {\em unrenormalized} boson self-energy at $T = 0$. In dimensional regularization, $\Pi_{\rm vac}(P^2) = \alpha b_0 \l[ \epsilon^{-1} - \ln(-P^2/\mu^2) \r] P^2$, where $b_0 = 4\pi\beta_0$ and $\beta_0$ is the leading coefficient of the $\beta$-function.
For a specific $P^2 = t_r$, the matrix element reads explicitly
\[
 \frac1{t_r}\, 
 \frac\alpha{1-\alpha b_0\l[ \epsilon^{-1}-\ln(-t_r/\mu^2) \r]}
 \equiv
 \frac{\alpha(t_r)}{t_r} \, .
\]
A measurement then specifies the {\em renormalized} coupling $\alpha(t_r)$ at the scale $t_r$ (as introduced on the rhs) which is related to the (infinite) bare coupling by
\be
 \alpha^{-1}(t_r) 
 = 
 \alpha^{-1} - b_0\l[ \epsilon^{-1} - \ln(-t_r/\mu^2) \r] .
 \label{eq: alpha bare vs run}
\ee
An equivalent relation holds for the coupling $\alpha(t)$ at an arbitrary scale $t$, consequently $\alpha^{-1}(t) = \alpha^{-1}(t_r) + b_0\, \ln(t/t_r)$ or, in a common alternative form,
\be
  \alpha(t) = \l[ b_0\, \ln(|t|/\Lambda^2) \r]^{-1} \, .
  \label{eq: alpha run}
\ee
It is underlined that the momentum dependence of the renormalized (`running') coupling is fully specified by its value at a certain scale $t_r$ or, equivalently, by the parameter $\Lambda$.

In a (thermal) medium, the boson self-energy has the generic structure
\[
  \Pi^i
  =
  \alpha b_0\l[
    \l( \epsilon^{-1} - \ln(-P^2/\mu^2) \r)P^2 + f^i(p_0,p)
  \r] ,
\]
where the finite `matter' contributions $f^i$ differentiate transverse and longitudinal modes ($i=t,l$).
Then, utilizing (\ref{eq: alpha bare vs run}), the in-medium scattering matrix can be rewritten in terms of the renormalized coupling,
\bea
	\frac\alpha{P^2-\Pi^i}
	&=&
	\frac{P^{-2}}
		{\alpha^{-1}-b_0[\epsilon^{-1}-\ln(-P^2/\mu^2)+f^i/P^2]}
	\nonumber \\
	&=&
	\frac{P^{-2}}{\alpha^{-1}(P^2) - b_0 f^i/P^2}
	= 
	\frac{\alpha(P^2)}{P^2 - \Pi_{\rm mat}^i} \, .
	\label{eq: M}
\eea
This distinct form, where the divergent vacuum contribution of the self-energy is `absorbed' in the running coupling, elucidates that the matrix element depends only on the physical parameter $\Lambda$ and the matter part of the self-energy, $\Pi_{\rm mat}^i = \alpha(P^2) b_0 f^i$ .
Thus the effective IR cut-off for the energy loss is, as expected,
related to the Debye mass \footnote{This is evident for the longitudinal contribution, while soft `magnetic' gluons are screened {\em dynamically} at the same scale.}.
The main emphasis here, though, is on the renormalized coupling in Eq.~(\ref{eq: M}) and, consequently, also in the resulting differential cross section: the scale of the running coupling is set by the virtuality $P^2 = t$. 

It is physically intuitive that this fact is generic. Thus, instead of a detailed analysis of loop corrections in QCD (which is more complex), it is useful to invoke a more comprehensive argument.
The vacuum differential cross section, as a quantity with an unambiguous normalization, obeys a Callan-Symanzik equation with $\gamma \equiv 0$ \cite{PeskinSchroeder}, 
\be
	\l[ M\, \frac\partial{\partial M}
	 + \beta(g)\, \frac\partial{\partial g} \r] \frac{d\sigma(t; M,g)}{dt}
	= 0 \, ,
\ee
where $g = \l( 4\pi\alpha(M^2) \r)^{1/2}$ is related to the coupling at a given renormalization point $M$. The general solution of this linear partial differential equation is a function $h(x)$, whose argument $x = g(M)$ satisfies $M dg/dM + \beta(g) = 0$. 
To investigate, in this line of thoughts, the dependence on the momentum scale $Q$, with $Q^2 = q^2 = t$, note that, in the limit of small $t$, the most general form of the cross section is $d\sigma/dt = {\cal S}(Q/M, g)/t^2$. Consequently, the function ${\cal S}$ obeys $[ -Q\, \partial_Q + \beta(g)\, \partial_g ] {\cal S} = 0$ (mind the minus sign). The corresponding characteristic equation, with $\beta(g) = -\beta_0 g^3$ at leading order, then leads readily to the running coupling as introduced above.
In other words, renormalization group invariance implies that loop corrections to the differential cross section can be `absorbed' in the tree level expression by replacing $\alpha \rightarrow \alpha(t)$.

A running coupling, as given by Eq.~(\ref{eq: alpha run}) with $b_0 = (11 - \frac23\, \nf)/(4\pi)$ in QCD, alters the integral (\ref{eq: Bjorkens t-int}) \footnote{Aiming here only at a leading order result, details related 
	to the broken Lorentz invariance in the presence of a medium can be 
	omitted.},
\bea 
	&&\!\!\!\! \Phi\int_{t_1}^{t_2}\! dt\, \frac{d\sigma_{\!js}}{dt}\, \omega
	\,=\,
	-\frac{\pi C_{\!js}}{k\, b_0^2}
	\int_{t_1}^{t_2} \frac{dt}{t\ln^2(|t|/\Lambda^2)}
	\nonumber \\
	&&=\,
	\frac{\pi C_{\!js}}{k\, b_0^2}
	\l. \frac1{\ln(|t|/\Lambda^2)} \r|_{t_1}^{t_2}
	\,=\,
	\frac{\pi C_{\!js}}{k\, b_0}
	\l[ \alpha(\mu^2) - \alpha(|t_1|) \r] . \rule{7mm}{0mm}
	\label{eq: t-int}
\eea
Opposed to the expression (\ref{eq: Bjorkens t-int}), the weighted cross section is UV finite -- due to the asymptotic weakening of the strong interaction, large-$t$ contributions are suppressed. For hard jets, the integral (\ref{eq: t-int}) becomes independent of the energy $E$, i.\,e., it is then controlled solely by the coupling at the screening scale. The necessary condition $|t_1| \sim TE \gg \mu^2 \sim m_D^2 \sim \alpha T^2$, is, for weak coupling, actually much less restrictive than the previous assumption $E \gg T$ to simplify the kinematics. 
Anticipating an extrapolation of the final result for $dE/dx$ to larger coupling, where the hierarchy of scales $\sqrt\alpha T \ll T$ breaks down, it is mentioned that $m_D \ksim 3T$ (see, e.\,g., \cite{Peshier:2006ah}). Hence from this perspective, the collisional loss could become $E$-independent already for jet energies exceeding several GeV's -- provided, of course, that the perturbative framework gives at least a semi-quantitative guidance at larger coupling, which will be advocated below.

The jet's collisional energy loss approaches, thus,
\be
	\frac{dE_j}{dx}
	=
	\pi \frac{\alpha(\mu^2)}{b_0} \sum_s C_{\!js}\int_{k^3} 
	\frac{\rho_s(k)}k \, . 
	\label{eq: dEdx before k-int}
\ee
In the ideal gas limit, the remaining integration yields
\be
	\frac{dE_{q,g}}{dx}
	=
	\l( \T\frac23 \r)^{\!\pm 1}
 	2\pi \l( 1+\T\frac16 \nf \r) \frac{\alpha(\mu^2)}{b_0}\, T^2 \, ,
 	\label{eq: dEdx}
\ee
which departs radically from previous expressions as Bjorken's (\ref{eq: dEdx Bjorken}). Aside from the modified cut-off dependence discussed above, the QCD collisional loss is proportional to $\alpha$ (instead of $\alpha^2$). It is highlighted that the considerations also show that the relevant scale for the coupling is the (perturbatively soft) screening mass rather than a `characteristic' thermal (hard) scale $\sim T$, as commonly presumed.

In order to quantitatively compare Eq.~(\ref{eq: dEdx}) to previous estimates it is necessary to specify parameters, namely $\Lambda$ in (\ref{eq: alpha run}) and the cut-off $\mu$, i.\,e., the Debye mass. 
Similar renormalization arguments as above lead in \cite{Peshier:2006ah} to an implicit equation for $m_D$; to leading order 
\be
	m_D^2 = \T\l( 1+\frac16 \nf \r) 4\pi\alpha(m_D^2)\, T^2 \, ,
	\label{eq: mD}
\ee
whose solution can be given in terms of Lambert's function.
Obviously, this mended perturbative formula is justified strictly only at temperatures $T \gg \Lambda$. Notwithstanding this, it is found in {\em quantitative} agreement with lattice QCD calculations down to $\tilde T \approx 1.2T_c$ \footnote{In contrast, the `conventional' expression, 
	$\sim \sqrt{\alpha(2\pi T)}\, T$, deviates from the lattice results by an almost constant factor of 1.5 in a large temperature range.}.

It may come as a surprise that a leading order formula reproduces non-perturbative results \footnote{As argued e.\,g.\ in \cite{Peshier:1998ei}, 
	perturbative expansions could be asymptotic series (at best), implying an 
	optimal truncation order {\em inversely} proportional to the coupling. Hence 
	for strong coupling, a leading order result might be more adequate than 
	sophisticated high-order calculations  -- somewhat opposed to naive 
	intuition.}.
Thus, it is worth emphasizing that the adjusted parameter $\Lambda = 205\,$MeV for $\nf = 2$ is right in the expected ballpark, enervating a possibility of an uninterpretable `overstrained' fit. 
Moreover, the 1-loop running coupling (\ref{eq: alpha run}) with the {\em same}
$\Lambda$ reproduces lattice calculations for another quantity, namely the QCD vacuum potential $V(r)$, in fact up to large distances corresponding to $\alpha(r^{-2}) \approx 1$ \cite{Peshier:2006ah}. 
Although at first sight rather different, the potential at $T = 0$ and the Debye mass in the medium are closely related by the renormalized $t$-channel scattering discussed above -- as is the collisional energy loss.
In other words (tidying the order of the arguments): one can renormalize the theory at $T=0$ (i.\,e., determine once and for all $\Lambda$ from $V(r)$), verify the applicability of the perturbative approach for larger couplings as relevant near $T_c$ by successfully calculating $m_D(T)$, and then make predictions for $dE/dx$.

\begin{figure}[ht]
 \centerline{\includegraphics[width=9cm]{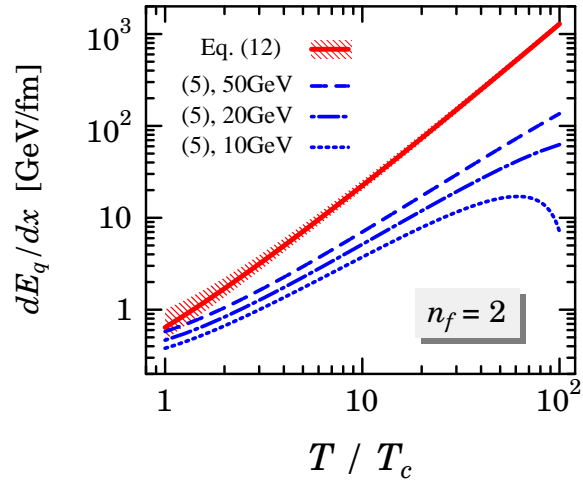}}
 \vspace*{-5mm}
 \caption{Light quark collisional energy loss: Eq.~(\ref{eq: dEdx}) vs.\ the 
	prevalent expression (\ref{eq: dEdx Bjorken}) (which yields negative values
	for very large $T$) for representative jet energies. For details see text.
 	\label{fig: dEdx}}
\end{figure}

With this justification, an extrapolation of Eq.~(\ref{eq: dEdx}) as presented in Fig.~\ref{fig: dEdx} might be not too unreasonable.
Assumed here is $T_c = 160\,$MeV and $\mu^2 = [\frac12,2] m_D^2$ to estimate the uncertainty due to the IR cut-off.
Shown for comparison are results from (\ref{eq: dEdx Bjorken}); here $\mu = m_D$ (likewise from (\ref{eq: mD})) albeit $\alpha$ in the prefactor (unjustified, but as often presumed) fixed at the scale $Q_T = 2\pi T$.
It turns out that already near $T_c$, the estimates from formula (\ref{eq: dEdx}) exceed those from (\ref{eq: dEdx Bjorken}), even for rather large values of $E$.

\begin{figure}[ht]
 \centerline{\includegraphics[width=9cm]{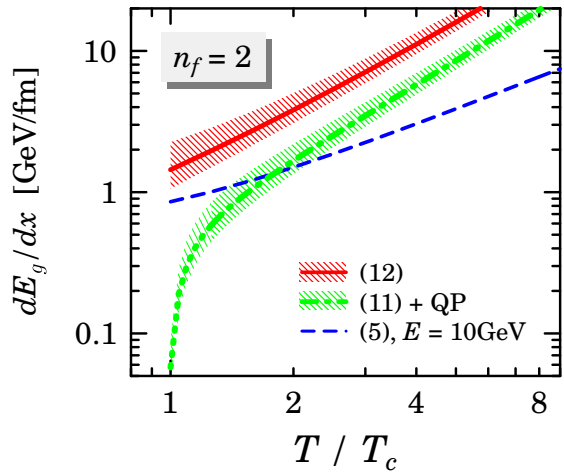}}
 \vspace*{-5mm}
 \caption{Influence of reduced scatterer density near $T_c$ on the gluon 
	energy loss. Below $\tilde T \approx 1.2T_c$, Eq.~(\ref{eq: mD}) 
	over-estimates the Debye mass as obtained in lattice QCD; hence $dE/dx$ 
	could be slightly larger than depicted by the dotted line.
 	\label{fig: dEdx extra}}
\end{figure}

For phenomenological implications it is instructive to take further into account a main effect of the strong interaction near $T_c$. From the distinct decrease of the QGP entropy seen in lattice QCD calculations \cite{Karsch:2000ps}, one can, on general grounds, infer a similar behavior for the number densities $\rho_s$. In the framework of the quasiparticle model \cite{Peshier:1999ww}, the ideal distribution functions in Eq.~(\ref{eq: dEdx before k-int}) are to be replaced by $d_s n_\pm(\sqrt{m_s^2(T)+k^2})$, where the effective coupling in the quasiparticle masses, $m_s^2 \propto \alpha_{\rm eff}(T) T^2$, is adjusted to lattice results for the entropy.
As shown in Fig.~\ref{fig: dEdx extra}, near $T_c$ -- despite the strong interaction -- the plasma becomes transparent due to the reduced number of `active' degrees of freedom.
Qualitatively, this characteristic behavior is in line with a rather sudden transition from small to large jet quenching when going from SPS to RHIC energies.

\smallskip

In conclusion, it has been demonstrated in the context of thermal field theory that renormalization does not only dictate the value of the coupling for a given quantity, but that the running of the coupling can also influence the parametric behavior of results.
For the QCD collisional energy loss, the relevant scale in $\alpha(t)$ is the (perturbatively soft) screening mass $m_D \sim \sqrt\alpha T$ (instead of $T$, as commonly presumed). The increasing coupling at soft momenta leads to a parametric enhancement compared to previous calculations, see Eqs.~(\ref{eq: dEdx}) vs.\ (\ref{eq: dEdx Bjorken}). On the other hand, due to asymptotic freedom, the collisional energy loss becomes independent of the jet energy $E$.
Thus, the asymptotic behavior of the underlying interaction makes the energy loss qualitatively different in QCD and QED (where an analog of Eq.~(\ref{eq: dEdx Bjorken}) indeed holds).

Except very near $T_c$, Eq.~(\ref{eq: dEdx}) suggests a larger collisional energy loss than previously estimated \cite{Bjorken:1982tu, Braaten:1991jj}. 
This finding can be interpreted as a facet of the `strongly coupled' QGP (sQGP), which is characterized by large cross sections.
In fact, $\sigma = \int dt\, d\sigma/dt$ with running coupling can actually be an order of magnitude larger than expected from the widely used expression $\sigma_{\alpha\, {\rm fix}} \propto \alpha^2(Q_T^2)/\mu^2$.
Thus, the present approach gives a consistent and simple explanation of  phenomenologically inferred $\sigma \sim {\cal O}(10)$mb \cite{Molnar:2001ux, Peshier:2005pp}.

Close to $T_c$, though, the particle density is known to be substantially reduced. This implies that, irrespective of the strong coupling, the sQGP becomes transparent near the (phase) transition.
Such a distinct temperature dependence of the energy loss should be observable. The quantification of this effect (including a discussion implications of the running coupling for the radiative energy loss) will be the subject of a forthcoming study.

\smallskip

\noindent
{\bf Acknowledgments:} I thank J.~Aichelin, W.~Cassing, S.~Jeon, M.~Thoma and in particular S.~Leupold and S.~Peign\'e for fertile discussions.
This work was supported by BMBF.

\end{document}